# Silanization Strategies for Tailoring Peptide Functionalization on Silicon Surfaces: Implications for Enhancing Stem Cell Adhesion

Melissa Kosovari[1,2,3], Thierry Buffeteau[5], Laurent Thomas[5], Andrée-Anne Guay Bégin[2,3], Luc Vellutini[5], James D. McGettrick[4], Gaétan Laroche[2,3,□,*], Marie-Christine Durrieu[1,□,*]

[1]Univ. Bordeaux, CNRS, Bordeaux INP, CBMN, UMR 5248, F-33600 Pessac, France.

[2]Laboratoire d'Ingénierie de Surface, Centre de Recherche sur les Matériaux Avancés, Département de Génie des Mines, de la Métallurgie et des Matériaux, Université Laval, 1065 Avenue de la médecine, Québec G1V 0A6, Canada

[3]Axe médecine régénératrice, Centre de Recherche du Centre Hospitalier Universitaire de Québec, Hôpital St-François d'Assise, 10 rue de l'Espinay, Québec G1L 3L5, Canada

[4]College of Engineering, Swansea University Bay Campus, Swansea SA1 8EN, United Kingdom

[5]Univ. Bordeaux, CNRS, Bordeaux INP, ISM, UMR 5255, F-33400 Talence, France

□ Equally contributed.

* Corresponding authors

## Abstract

Biomaterial surface engineering and integrating cell-adhesive ligands are crucial in biological research and biotechnological applications. The interplay between cells and their microenvironment, influenced by chemical and physical cues, impacts cellular behavior. Surface modification of biomaterials profoundly affects cellular responses, especially at the cell-surface interface. This work focuses on enhancing cellular activities through material manipulation, emphasizing silanization for further functionalization with bioactive molecules like RGD peptides to improve cell adhesion. The grafting of three distinct silanes onto silicon wafers using both spin coating and immersion methods was investigated. This study sheds light on the effects of different alkyl chain lengths and protecting groups on cellular behavior, providing valuable insights into optimizing silane-based self-assembled monolayers (SAMs) before peptide or protein grafting for the first time. Specifically, it challenges the common use of APTES molecules in this context. These findings advance our understanding of surface modification strategies, paving the way for tailoring biomaterial surfaces to modulate cellular behavior for diverse biotechnological applications.

Email addresses: gaetan.laroche@gmn.ulaval.ca (Gaétan Laroche), marie-christine.durrieu@inserm.fr (Marie-Christine Durrieu)

## 1. Introduction

The field of biomaterial surface engineering and cell-adhesive ligands is expanding, representing a key area of biological investigation and a promising avenue for biotechnological applications[1]. In their natural environment, cells encounter different chemical and physical factors on various scales, from nanometers to hundreds of microns[2,3].

Cellular behaviors are heavily influenced by the microenvironment in which they exist[4]. Indeed, cells can sense and respond to various physical and functional properties of their external environment through the formation of adhesions, a critical process for their decision-making and behavior[5,6]. Cell membrane receptors, like integrins, play a significant role in transmitting physical information from the microenvironment into intracellular signaling pathways, ultimately leading to changes in cell proliferation, differentiation, migration, or apoptosis behaviors[7,8]. Surface modification significantly affects the modulation of cellular activities because all functions are concentrated at the cell-surface interface[9]. Several strategies to control the biochemical properties of materials used in tissue engineering mainly influence focal adhesion (FA) formation and growth[10]. One of the most common is functionalization with ligands. However, various proteins or peptides involved in cell adhesion may adsorb unevenly or not be readily available to cells, thus impairing FA dynamics[11,12]. Techniques used for chemical modification of a biomaterial aim to add a functional group, such as amines, carboxylic acids, or thiols, to the surface[13]. These groups allow for the covalent immobilization of polypeptides, whole proteins, glycosaminoglycans, or other selected molecules that can promote cell adhesion[14]. Creating robust and functional organic films as primary layers with controlled density is crucial in the field of biotechnological applications ranging from biosensors to tissue engineering. Silane-based self-assembled monolayers (SAMs) find diverse applications on silica substrates, leading to extensive research spanning several decades to refine fabrication techniques[15]. SAMs have been widely used to generate specific functional groups on oxide surfaces. The robustness of trialkoxysilanes-based SAMs comes from forming a siloxane bonds network by the horizontal polymerization between neighboring silanes and the free hydroxide groups on the silica surface[11].

Amino-terminated SAMs have been widely used for covalently immobilizing biomolecules to control cell behavior. In the intricate extracellular matrix (ECM) environment, where cells respond to dynamic cues, there is a crucial need to replicate the native microenvironment. This replication is designed to facilitate the migration of stem cells while concurrently enhancing the microenvironment's capacity to attract cells on the surface. Conversely, when the surface remains untreated, immune cells often identify it as foreign, triggering an exaggerated inflammatory response[16]. When attaching organic molecules to inorganic surfaces, one common practice involves depositing self-assembled monolayers to modify these surfaces with reactive groups, such as amino groups achieved through the coupling of organosilanes. Subsequently, biomolecules can bind to these chemically altered surfaces by reacting with the introduced groups[17]. Chuah et al. observed significant transformations in the native surface properties of PDMS (polydimethylsiloxane), with a notable improvement in MSC (mesenchymal stem cells) adhesion evident upon APTES (3-aminopropyl-triethoxysilane) modification. This highlights the profound influence of silanization on PDMS surface characteristics, including decreased hydrophobicity, increased protein immobilization, and variations in nanotopography[18]. Furthermore, even the amine groups on APTMS-modified (3-aminopropyl-trimethoxysilane) surfaces exhibited a more efficient improvement in cell adhesion compared to the hydroxyl groups found on oxygen plasma-treated PDMS surfaces. The absorption of proteins emerges as an early in vivo occurrence in the interplay between implanted biomaterial and living tissue. Consequently, a surface featuring an organized arrangement of functional groups can serve as a conducive site for cell growth in the development of biomaterials[19]. The biological properties of silanized graphene oxide (SiGO) were explored in the context of material biomedical applications. Silanization using APTES enhanced the material in vitro cytocompatibility and mitigated its in vivo toxicity. The findings additionally indicated that SiGO exhibits superior biocompatibility compared to graphene oxide (GO), suggesting potential applications in the field of biomedical engineering [20].

Amino-terminated SAMs are generally prepared using commercially available organosilanes such as 3-aminopropyl-triethoxysilane (APTES) and 11-aminoundecyltriethoxysilane (AUTES). Silanization usually proceeds by immersing the material in an organosilane solution using an organic solvent. This so-called immersion silanization method is simple and easy to conduct. However, controlling the layer structure of amino-terminated SAMs is difficult, leading to the non-reproducibility of surface functionalization[21,22]. The multiple possibilities of interactions of the amino-end group by hydrogen bonds and electrostatic attractions render the molecular organization in the organic layer difficult and, therefore, affect the accessibility and reactivity of the amine group[23]. A terminal amino protecting group such as N-phthalimide can be used to avoid multiple interactions[24] to prevent uncontrolled orientation of the amino silanes. Two prevalent methods for surface modification involve silanes: solution-based reactions and vapor-phase deposition. The traditional solution immersion technique submerges clean substrates in a diluted solution of silylated coupling agents within a controlled environment. However, it generates significant chemical waste due to the use of organic solvents. In contrast, the chemical vapor deposition (CVD) method exposes clean substrates to silane vapor, eliminating the need for organic solvents but requiring ultra-high vacuum conditions and elevated temperatures. Less commonly used techniques, such as the Langmuir−Blodgett method and spin coating, have seen limited application and less extensive study than the primary methods[25,26]. Here, we propose to employ spin coating for the first time as a methodology for rapidly and cost-effectively immobilizing silane molecules on silicon surfaces for bioactive molecule grafting. According to the literature, the fabrication of silane-based SAMs utilizing the spin coating technique was explored as an alternative, adaptable, and cost-effective approach in contrast to the traditional solution immersion method[27].

This study aimed to assess the influence of diverse silanization protocols (using conventional immersion method *vs* spin coating) using heterobifunctional silanes to obtain amino-functionalized silicon wafer surfaces for further RGD peptide conjugation and their impact on mesenchymal stem cell adhesion[12]. Organosilane molecules with identical trialkoxy anchoring groups but different alkyl chain lengths were investigated. On the other hand, a comparative analysis included molecules of identical chain lengths with or without N-phthalimide protecting groups opposite to the surface. This exploration aimed to evaluate the influence of the different alkyl chain lengths and the protecting group presence on the silane-based monolayer's distribution on the silicon surface, as well as its implications for the accessibility and subsequent efficacy of peptide grafting. This investigation enabled to highlight differences in surface functionalization using different silanization methods and their subsequent influence on cellular behavior, mainly focusing on adhesion dynamics and cell spreading tendencies. This study constitutes the first step toward the investigation of stem cell differentiation using fully optimized and controlled surface chemistries for further grafting of various bioactive molecules.

## 2. Materials and Methods

### Materials

Silicon wafers (100 mm, thickness 500-550 µm) were obtained from Pure Wafer, San Jose, CA. The $SiO_2$/Au substrates used for PM-IRRAS experiments, specifically Goldflex mirrors with a $SiO_2$ protective layer (Goldflex PRO, reference 200785), were provided by Optics Balzers AG. Hydrogen peroxide (33 wt %), sulfuric acid (concentrated), acetone, ethanol, anhydrous toluene, dimethylsulfoxide (DMSO), 3-aminopropyltriethoxysilane (APTES), and succinimidyl-4-(p-maleimidophenyl) butyrate (SMPB) were all acquired from Sigma-Aldrich, France, while 11-aminoundecyltriethoxysilane (AUTES) was purchased from Gelest, Inc., Morrisville, PA. Alk-Phtha was synthesized in the Molecular Chemistry & Materials team at ISM, Bordeaux University, under the supervision of Prof. Luc Vellutini. GeneCust in Boynes, France, synthesized the fluorophore-tagged peptide CG-K(PEG3-TAMRA)-GGRGDS, referred to as RGD (molecular weight = 1437 g $mol^{-1}$). Bone marrow mesenchymal stem cells (hBMSCs), mesenchymal stem cell growth medium 2, and SupplementMix MSC growth medium 2 were procured from PromoCell GmbH, located in Heidelberg, Germany. Monoclonal Anti-Vinculin antibody (produced in mice) was obtained from Sigma-Aldrich, St. Louis, USA. Dulbecco’s modified Eagle’s medium (DMEM), Dulbecco’s phosphate buffered saline (1×) (PBS), Trypsin-EDTA, fetal bovine serum (FBS), Alexa Fluoroshield™ 488 phalloidin, secondary antibodies (goat anti-mouse lgG (H + L) highly cross-adsorbed secondary antibody Alexa Fluoroshield™ 568 and DAPI were procured from Thermo Fisher Scientific (USA). Antibiotic/antimycotic solutions were obtained from GE Healthcare Life Sciences (USA).

### Synthesis of phthalimide-terminated coupling agents

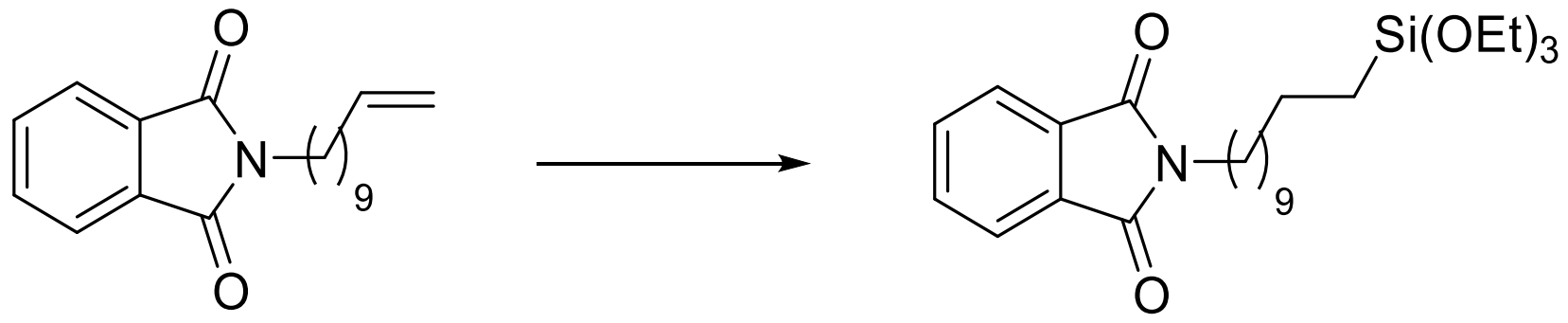

**1** **2**

**Scheme 1.** Synthesis of phthalimide-terminated silylated coupling agent **2**: compound **1** was synthesized according to ref 17. Compound **2** was synthesized by a hydrosilylation reaction using $HSi(OEt)_3$ (5 eq.), MeCN, cat. Karstedt (0.0125 eq. Pt), 60 °C, 1 h.

The synthesis of coupling agents terminated with phthalimide (**2**) was performed to combine the same structure as AUTES but with the end amino group protected by a N-phthalimide group until deprotection and further grafting (Alk-$NH_2$). Phthalimide-terminated triethoxysilane **2** was synthesized by using the platinum-catalyzed hydrosilylation of alkene precursor **1**, as shown in Scheme 1. A dry Schlenk tube under inert atmosphere was filled with the alkene precursor **1** (1.014g, 3.4 mmol, 1eq) and 30 mL of anhydrous acetonitrile when needed. Triethoxysilane (2.33g, 14.2 mmol, 4 eq) was introduced in the Schlenk tube followed by Karstedt's catalyst (326 µl, 1 mmol, 0.3 eq). After 2 h under stirring at 60 °C, evaporation of the solvent occurred. The brown crude product underwent filtration on a silica pad with $CH_2Cl_2$, resulting in the acquisition of a colorless oil of compound **2** with an 85% yield after evaporation of $CH_2Cl_2$.

**$^1$H NMR** ($CDCl_3$, 300 MHz) δ (ppm): 7.86 (dd, J = 5.4, 3.0 Hz, 2H, 2 $CH_{ar}$), 7.72 (dd, J = 5.5, 3.1 Hz, 2H, $CH_{ar}$), 3.83 (q, 6H, 3 $CH_2OSi$), 3.70 (t, *J* = 7.2 Hz, 2H, $CH_2N$), 1.34 - 1.26 (m, 18H, 9 $CH_2$), 1.25 (t, 9H, 3 $CH_3$), 0.65 (m, 2H, $CH_2Si$).

**$^{13}$C NMR** ($CDCl_3$, 100 MHz) δ (ppm): 168.6 (2 CO), 134.0 (2 $CH_{ar}$), 132.3 (2 $C_{ar}$), 123.3 (2 $CH_{ar}$), 58.4 (3 $CH_2OSi$), 38.2 ($CH_2N$), 33.3 ($CH_2$), 29.7 ($CH_2$), 29.6 (2 $CH_2$), 29.4 ($CH_2$), 29.3 ($CH_2$), 28.7 ($CH_2$), 27.0 ($CH_2$), 22.7 ($CH_2$), 18.44 ($CH_3$), 10.5 ($CH_2Si$).

**$^{29}$Si NMR** ($CDCl_3$, 79 MHz) δ (ppm): -41.2. **HRMS** (ESI)**:** calculated for $C_{25}H_{41}O_5NSiNa^+$ ([M+$Na^+$]): 486.2646, found 486.2640.

**Sample preparation**

*Activation.* Silicon wafers were resized into appropriate dimensions, approximately 0.5 cm × 0.5 cm. The silicon slides underwent ultrasonic cleansing through successive immersions in acetone (10 min), deionized water (10 min), and methanol (10 min). Following this, the slides were activated by further immersion in a piranha solution (mixture of 10 mL of $H_2O_2$ (33% wt.) and 30 mL of concentrated $H_2SO_4$) for 15 min in an ultrasonic bath. Subsequently, the samples were subjected to three rounds of ultrasonic cleansing in water for 10 minutes and in methanol for 2 minutes. $SiO_2$/Au substrates were resized into suitable dimensions, approximately 2 cm × 2 cm. These substrates underwent thorough washing with Milli-Q water (18 MΩ cm), followed by a 15-minute sonication in chloroform and exposure to UV-ozone (185-254 nm) for 30 minutes. They were promptly utilized for silanization via either the spin-coating process or the immersion solution method.

*Silanization by spin coating and immersion method.* Two silanization techniques, spin coating and immersion, were investigated for efficient silanization, with three different silane molecules (Scheme 2): APTES (3-aminopropyltiethoxysilane) and AUTES (11-aminoundecyltriethoxysilane) commercial molecules, as well as a third silane molecule, Alk-Phtha (phthalimide-terminated undecyltriethoxysilane). The time of exposure to the silane solution was optimized according to Rouvière et al[24].

*Silanization by immersion.* After silicon surface activation, the cleaned surfaces were immediately functionalized with silane molecules. The specimens were soaked in a 2% v/v silane solution diluted in anhydrous toluene for a duration of 1 hour. The final volume of the silane solution ranges between 40 and 60 ml to adequately cover all surfaces of the immersed samples. The process was carried out with stirring and in an inert nitrogen environment. Following silanization, the substrates underwent outgassing at 120°C under vacuum for 1 h. Following this, the samples were treated to ultrasonic cleaning with anhydrous toluene, deionized water and methanol (2 sessions of 5 minutes each) to eliminate any unbound molecules.

*Silanization by spin coating.* Silane molecules were dissolved in absolute ethanol under an inert nitrogen atmosphere to achieve a final concentration of 2% v/v. 160 µl of the silane solution prepared just before was utilized to coat the silicon or $SiO_2$/Au substrates (2 $cm^2$ dimensions) at a rotation of 6000 rpm for 60 s. The samples underwent overnight drying at 120 °C under vacuum. Subsequently, they were subjected to sonication in ethanol, water and methanol (2 cycles of 5 minutes each) for washing.

*SMPB Crosslinker conjugation.* Before peptide grafting, aminated silicon surfaces underwent conjugation with the heterobifunctional SMPB crosslinker. The succinimidyl functionality of this compound reacted with the surface amine groups, resulting in pendent maleimidyl moieties. These moieties subsequently reacted with thiol groups from the terminal cysteine of selected peptides. Initially, aminated surfaces were immersed in a solution of SMPB dissolved in DMSO at a concentration of 3 mg/mL for 2 hours in the dark. Following this, the substrates underwent a gentle washing with methanol and were rinsed in DMSO for 10 minutes using an ultrasonic bath. After the cleaning process, the substrates were air-dried and kept in the dark for 24 hours before peptide grafting.

*RGD peptide grafting*. SMPB-grafted surfaces were immersed in 300 µL of 0.1 mM CRGD-TAMRA PBS1X solution in a dark, moist chamber at room temperature to achieve anchoring of fluorescently labeled RGD mimetic peptides (Scheme 3). Samples were then cleaned by ultrasonication six times for 15 min in nano-pure water to remove all non-covalently attached peptides. All reactions were performed under agitation at room temperature.

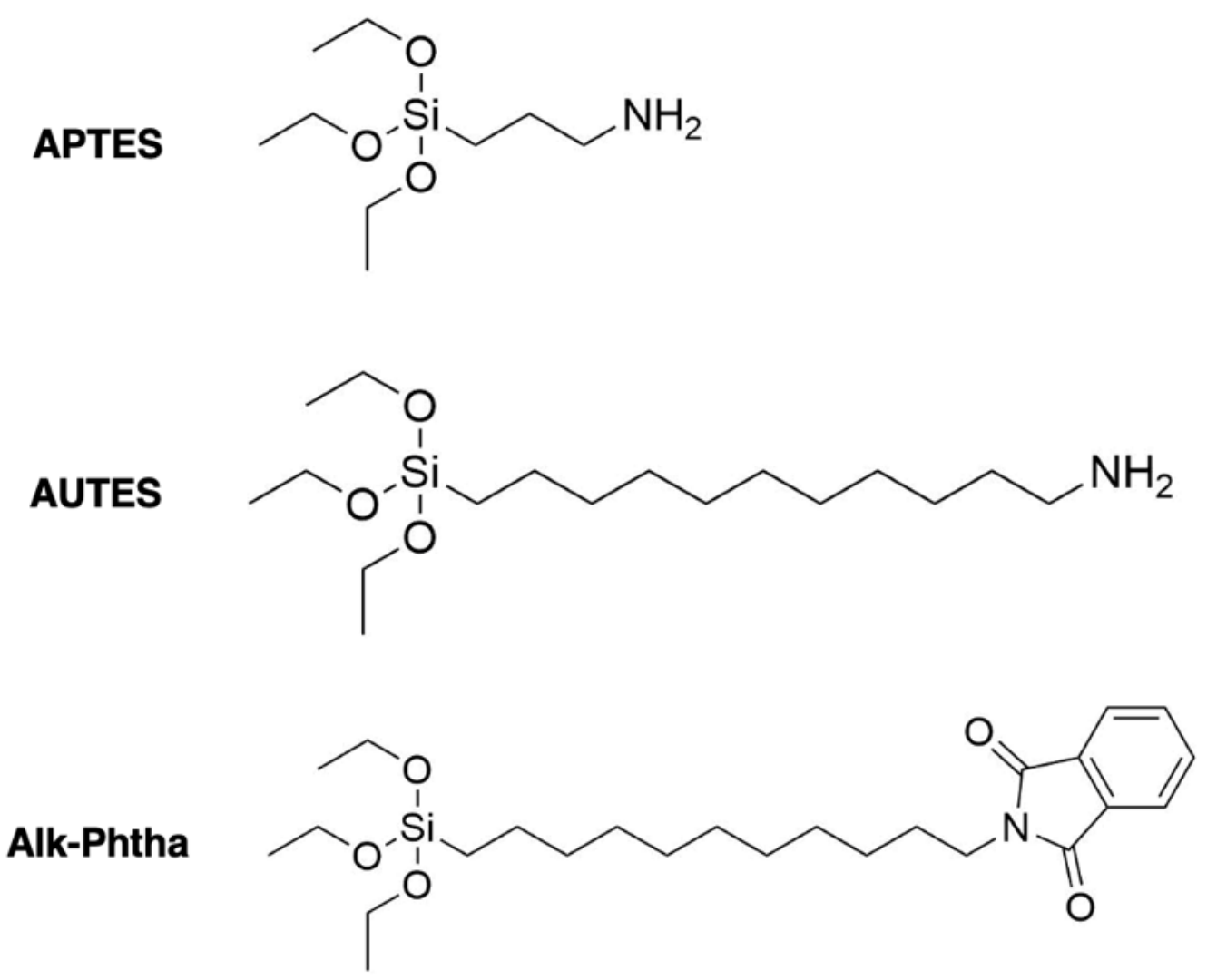

**Scheme 2**. Molecular structures of different silane molecules involved in the silanization process.

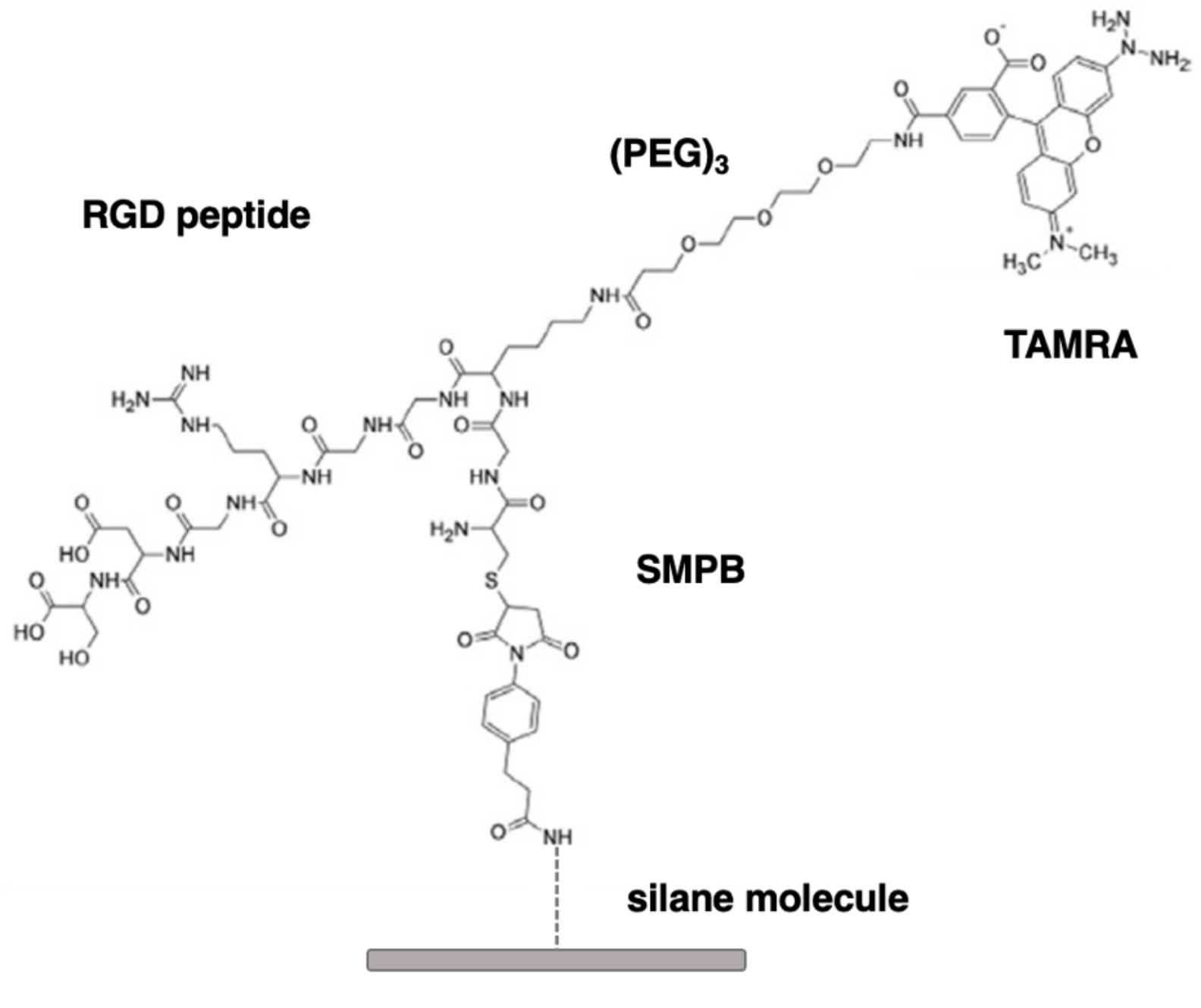

**Scheme 3**. Molecular structures of fluorophore labelled peptide anchored to SMPB crosslinker. [28]Adapted with permission from Hoesli C. et al.[28]

## Surface Characterization

*Water contact angle (WCA*). The assessment of surface wettability involved measuring water contact angles after the deposition of nano pure water droplets of 3 µL at 20 °C in static mode on a Krüss DSA 100 goniometer. The reported contact angle values resulted from the mean of three drops on different surfaces (n between 2 and 6).

*Atomic Force Microscopy (AFM).* The measurements were performed on a Bruker Dimension Fastscan AFM in tapping mode in air using a Bruker SNL-C cantilever (resonant frequency around 56 kHz and nominal curvature radius of 2 nm) (Bruker, Billerica, USA). AFM images treatment and roughness analysis were performed with Gwyddion software.

*X-Ray Photoelectron Spectroscopy (XPS).* XPS analysis was performed on an Axis Supra instrument (Kratos Analytical, UK) using monochromated Al Kα x-rays at 15 mA, 225 W emission, with the electron lenses set in the 'Hybrid' mode. For every sample. ann initial survey scan of pass energy 160 eV was performed. High-resolution spectra were then obtained for each element identified in the survey scan with a pass energy of 40 eV, step size of 0.1 eV, and a minimum dwell time of 500 ms. Where atomic concentrations are low for a given element, extra sweeps and longer dwell times were introduced to improve signal-to-noise ratio. The integral charge neutralizer was used to limit differential charging. Analysis of the data was conducted utilizing CasaXPS (Version 2.3.24PR1.0, Casa Software Ltd, UK). Spectra were fitted with a Shirley baseline, and a synthetic component was used to model the data. The synthetic components utilized the mixed Gaussian-Lorentzian (GL (30)) line shape, and FWHM was kept constant for a given element. The spectra were charge-corrected to correct the C(1s) hydrocarbon component to 285.0 eV. The area of each peak was corrected using the Kratos sensitivity factor library. The expected theoretical values were calculated only based on the atomic composition of the silane molecules, without considering the composition of the surface on which they are anchored.

*Polarization Modulation Infrared Reflection-Absorption Spectroscopy (PM-IRRAS).* PM-IRRAS experiments utilized a ThermoNicolet Nexus 670 FTIR spectrometer equipped with a PM-IRRAS optical bench[28]. Spectra were acquired with a resolution of 4 $cm^{-1}$ over a 5-hour acquisition period. Calibration was adjusted to present the PM-IRRAS spectra in IRRAS units[29,30]. Data collection occurred in a dry-air atmosphere following a 1-hour incubation in the chamber.

## Cell culture experiments

Commercially available Human Bone Marrow Mesenchymal Stem Cells (hBMSCs) were grown on culture flasks in DMEM medium enriched with 1% Pen Strep in serum-free conditions, subcultured using trypsin/EDTA 1x and sustained in a humidified atmosphere with 5% $CO_2$ at 37 °C. RGD-grafted 1 × 1 cm silicon substrates were placed in 24-well cell culture plates. A density of 3000 cells/$cm^2$ of hBMSCs at the third passage were seeded onto surfaces grafted all with RGD but treated with different silanization methods. The cells were cultured for 24 h in serum-free DMEM medium. This condition enables interactions between grafted RGD peptides and their cell surface receptors without interference with serum proteins.

*Immunocytochemical analysis*. Following 24 h of cell culture, hBMSCs underwent PBS 1X rinsing and fixation in 4% paraformaldehyde at 4 °C for 20 min. Samples were then permeabilized with 0.5% Triton X-100 in PBS 1X for 5 min and blocked with 1% Bovine Serum Albumin (BSA) in PBS 1X for 30 min at 37 °C. Incubation with 1:200 primary antibodies mouse monoclonal anti-vinculin in 1% BSA in PBS was carried out for 1 h at room temperature. Next, samples were washed in PBS 1X, prior incubation with the secondary antibody Alexa Fluor 568 goat anti-mouse IgG (H + L) (1:200 dilution) for 30 min at room temperature. Incubation with Alexa Fluor 488 phalloidin (1:40 dilution) for 1 h at 37 °C was performed in order to detect the presence of cytoskeletal fibers of F-actin. Three rinsing cycles with 1X PBS and subsequent labeling of the cell nuclei with 20 ug/ml DAPI for 10 min at room temperature followed. Finally, surfaces were examined using a Leica DM5500B microscope equipped with a motorized and programmable stage and a CoolSnap HQ camera controlled by Metamorph 7.6. Image J freeware (NIH, http://rsb.info.nih.gov/ij) was used for quantification of vinculin and phalloidin expression. All stained marker images were acquired at the same exposure time, using a 40x objective. Image files were opened with Image J and converted to 16-bit files. The intensity detection of the green and red colour emitted by the label, was obtained taking into consideration the subtraction of the background signal from the RGD-TAMRA surfaces. The values of fluorescence intensity were obtained from the analyses performed on at least 80 cells for each surface type.

## Statistics

Results are expressed as means ± SEM. Data were analyzed with Prism Software (GraphPad Software Inc.). Comparisons were made using ANOVA coupled with Tukey's multiple comparisons test. The statistical significance was defined for a P value of at least $< 0.01$.

## 3. Results

### Characterization of the biochemical modification of silicon surfaces

Silanization experiments were performed for all three molecules (APTES, AUTES, Alk-Phtha) using both methodologies (spin coating and immersion). From these experiments, we decided to select the best technique for each silane molecule in terms of reproducibility based on XPS analyses (data not shown), with five experiments conducted for each method. It turned out that spin coating was selected for APTES and Alk-Phtha silanization, while the immersion method was chosen for AUTES silanization. Accordingly, we aimed to compare the three most efficient strategies to choose the best one above all, which could stand as the optimal solution for silanization processes for biological purposes. Of note, Rouvière et al.[27], have shown by PM-IRRAS experiments that the grafting of Alk-Phtha by spin coating and immersion methods give the same results by optimizing the experimental conditions of deposition.

### Water Contact Angle and Atomic Force Microscopy

This technique provides a first indication that a change in the hydrophilic nature of the surface has occurred. Indeed, contact angle measurements made it possible to determine the hydrophobicity and hydrophilicity properties of the surface after each step of surface modification. After activation with piranha, the surface shows characteristic hydrophilicity (< 10°) due to the hydroxyl groups provided by the treatment. Following silanization, the increase in contact angle measurement confirms that wettability change has occurred, and the AUTES exhibited the highest hydrophobic character (73 ± 2°) as compared to the APTES surface (62 ± 9°) and deprotected Alk-$NH_2$ (67 ± 3°). The observed transition from high hydrophilicity after piranha treatment to increased hydrophobicity underscores the effectiveness of silanization surface modification. AFM measurements were carried out to determine surface roughness and complement the evaluation of hydrophilicity. These data show that the roughness of the various silane-conjugated surfaces ranges between 2.3 nm and 4.1 nm, meaning that the hydrophilicity differences between all investigated samples should be related to the chemical nature of the silane molecules as opposed to the surface roughness. These roughness measurements, as measured by AFM, are now provided along with the contact angle data in Table 1.

Table 1. Root mean square roughness (R rms) calculated by AFM measurements.

| **Substrates** | **Water Contact Angle** | **Roughness RMS (nm)** |
|---|---|---|
| Silicon | < 10° | 0.3 |
| APTES | 62 ± 9° | 3.4 |
| AUTES | 73 ± 2° | 2.3 |
| Alk-$NH_2$ | 67 ± 3° | 4.1 |

**X-Ray Photoelectron Spectroscopy**

The efficiency of the silanization reaction with the three different silane molecules was confirmed by XPS analysis. Each surface exhibited the expected elements, with an increase in the percentage of carbon recorded and a decrease in oxygen and silicon, as expected from the chemical structure of APTES, AUTES, and Alk-$NH_2$ (Table 2). Contamination from carbonaceous species is always detected on any surface and cannot be removed entirely, even in an ultra-high vacuum. The appearance of nitrogen clearly showed the efficiency of aminosilanization since no nitrogen was observed on the piranha-activated surface (Table 2).

A high presence of silicon was observed when compared with the chemical structure of the molecules (scheme 1) under all conditions. This can be attributed to the silicon wafer to which these molecules are attached. In fact, chemical composition analysis with XPS also detects the substrate-derived silicon component and not only that of the grafted silanes.

**Table 2**. XPS survey analyses of silicon surfaces after activation, silanization, and RGD grafting.

| | % C | % O | % Si | % N | N/C |
|---|---|---|---|---|---|
| Piranha | 13 ± 1 | 38 ± 2 | 49 ± 1 | | |
| APTES | 34 ± 2 | 31.2 ± 0.9 | 31 ± 2 | 4.2 ± 0.9 | 0.12 |
| RGD | 35 ± 2 | 28 ± 1 | 29 ± 1 | 5.2 ± 0.4 | 0.15 |
| AUTES | 65 ± 1 | 20 ± 1 | 12.2 ± 0.2 | 3.5 ± 0.3 | 0.05 |
| RGD | 55.2 ± 0.8 | 20.2 ± 0.3 | 14.7 ± 0.9 | 5.6 ± 0.2 | 0.1 |
| Alk-$NH_2$ | 17.6 ± 0.2 | 32.9 ± 0.2 | 48.2 ± 0.1 | 1.0 ± 0.1 | 0.06 |
| RGD | 20 ± 3 | 32 ± 1 | 43 ± 2 | 2.2 ± 0.6 | 0.11 |

Figure 1 (left) represents XPS high-resolution spectra of silanized surfaces in the Si2p region for the three different silane molecules. The Si2p profile of the APTES-modified surface is identical to the one observed for the Alk-$NH_2$-modified surface. In contrast, the Si2p XPS spectrum obtained after the AUTES silanization is quite different from those of APTES and Alk-$NH_2$-modified surfaces, especially for AUTES and Alk-$NH_2$, which display the same molecular structure (Scheme 1). The Si2p XPS signal globally shows two distinct doublets, corresponding to bulk silicon (99.1 eV) and silicon bonded to oxygen (103.8 eV). On the APTES and Alk-$NH_2$-modified surfaces, the ratio of the Si peak to the $SiO_2$ peak is approximately 70/30, but for the AUTES-modified surface, the ratio is 10/90. This increase in the contribution of $SiO_2$ on the AUTES-modified Si surface may be due to two factors that are consistent with a higher surface loading of AUTES relative to APTES and Alk-$NH_2$-modified surfaces. First, the addition of AUTES on the surface could result in the deposition of a thicker layer on top of the silicon than the deposition of the other two silane compounds, thus leading to an attenuation of the XPS signal of the underlying bulk Si (elemental) and increasing the relative contribution of both the native $SiO_2$ oxide on the wafer and of the silane functional group. Second, the AUTES is grafted more densely on the surface, decreasing the relative XPS signal of the underlying Si (elemental).

The C1s high-resolution spectra after RGD grafting on the different silanized surfaces are shown in Figure 1 (right). All spectra were fitted to 3 peaks having different components assigned to C-C/C-H at 285 eV, C-O/C-N at 286.43 eV, and the peptide signature of N-C=O and N-C(=O)-O at 288.52 eV. These results agreed with the literature for peptide XPS spectra, revealing that a successful peptide grafting was achieved on all silanized surfaces[31–33].

Table 2 demonstrates varying presence of amine functions depending on the silanization method, but the amount of amine in the surface region sampled by XPS (<10 nm depth) may not correspond directly to the amines being available for surface functionalization with the RGD. All three silanized surfaces showed increased nitrogen consistent with the increased proportion of nitrogen present in the RGD. The single sulfur atom present in the SMPB linking arm was not observed, XPS has a relatively low sensitivity of ~0.1 At%, so the relatively low number of sulfur atoms likely render sulfur below the detection limit. However, the SMPB linker arm and TAMRA fluorophore both contain a significant number of sp2 hybridized carbon atoms. The sp2 and sp3 C(1s) photoelectron peaks would be expected to overlap, but sp2 hybridized carbon atoms also display a distinct shake-up peak due to electron $\pi \rightarrow \pi^*$ transitions[34]. In this case a small feature corresponding to possible presence of the $\pi \rightarrow \pi^*$ is observed for only one of the RGD-modified films: Alk-$NH_2$.

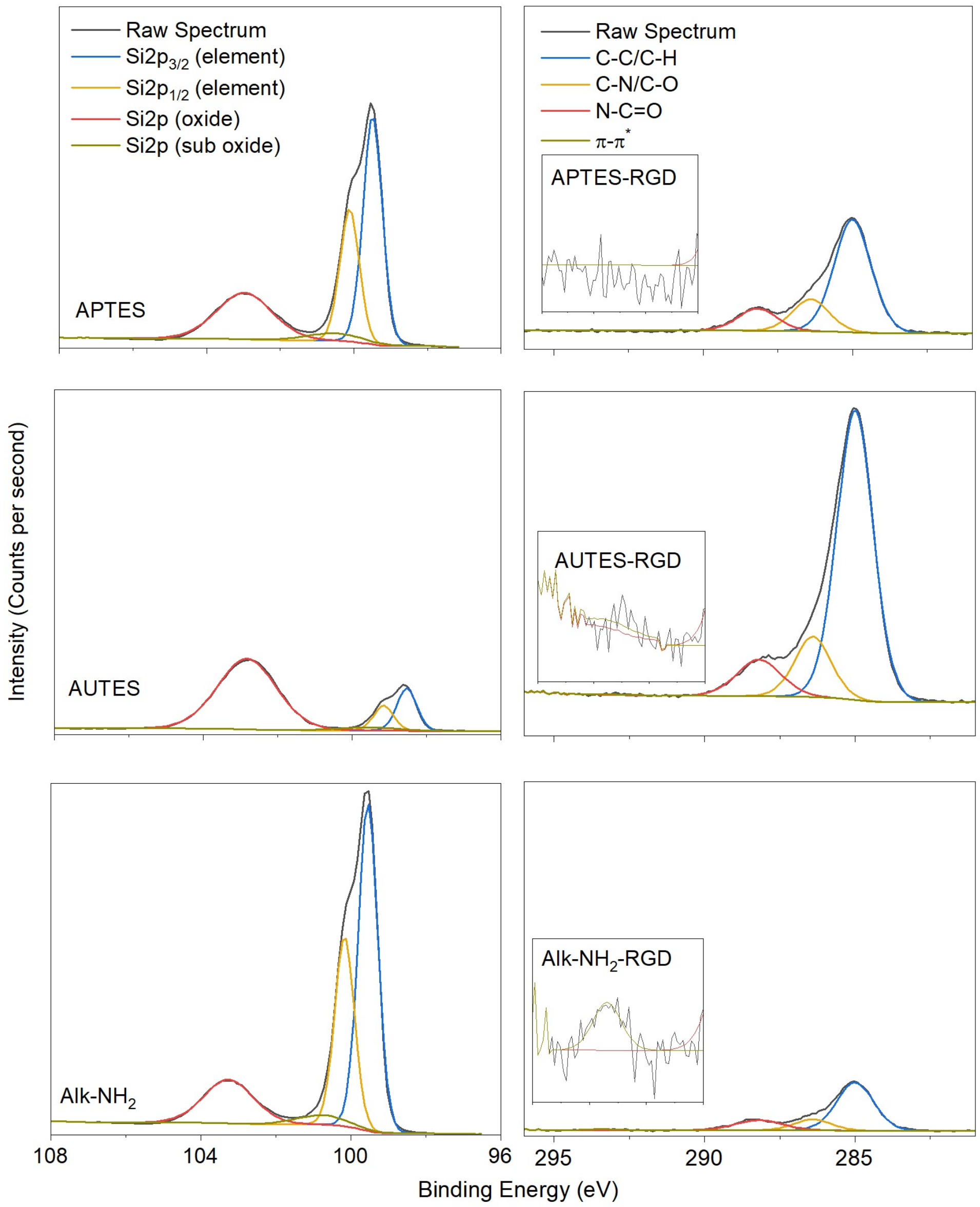

**Figure 1**. XPS high-resolution spectra of Si2p after silanization with APTES, AUTES, and Alk-$NH_2$ (left) and C1s after RGD grafting (right). Silanization of APTES and Alk-Phtha was achieved through spin coating using a 2% v/v solution in absolute ethanol, while AUTES silanization was conducted via immersion in a 2% v/v solution in anhydrous toluene.

## Polarization Modulation Infrared Reflection-Absorption Spectroscopy (PM-IRRAS)

Silane-based self-assembled monolayers (SAMs) are fabricated by chemically attaching silane layers that offer a prime opportunity for probing interactions at interfaces[35]. After their deposition, the self-assembled monolayers of silane molecules on the surface were also characterized by PM-IRRAS. The PM-IRRAS spectra of amine or phthalimide-terminated silane layers, prepared via spin coating (APTES, Alk-Phtha) and immersion (AUTES) are presented in Figure 2. The effectiveness of the grafting process was confirmed for all compounds by identifying distinct IR bands on the PM-IRRAS spectra (black spectra in Figure 2). First, the disorderly nature of the alkyl chains was revealed by the positioning of the antisymmetric ($\nu_a CH_2$) and symmetric ($\nu_s CH_2$) stretching vibrations of the methylene groups at 2927 and 2855 cm$^{-1}$, respectively. This aligns with the findings in existing literature[27,36]. From the intensity of the $\nu_a CH_2$ feature (0.0025), it can be deduced that the thickness of the Alk-Phtha layer is 15 Å, which is in agreement with the presence of a single monolayer[26]. For APTES and AUTES layers, the broad peak around 1625 cm$^{-1}$ was assigned to the bending mode ($\delta NH_2$) of the amino groups. The conversion of phthalimide groups into amine groups was performed by immersion in a methylamine solution.[33] This deprotection was confirmed in Figure 2c, by the complete disappearance of the bands at 1773 and 1715 cm$^{-1}$, assigned respectively to the in-phase and out-of-phase carbonyl stretching vibrations of the phthalimide terminal. Additionally, a new band appeared at 1625 cm$^{-1}$ due to the bending vibration of $NH_2$ groups.

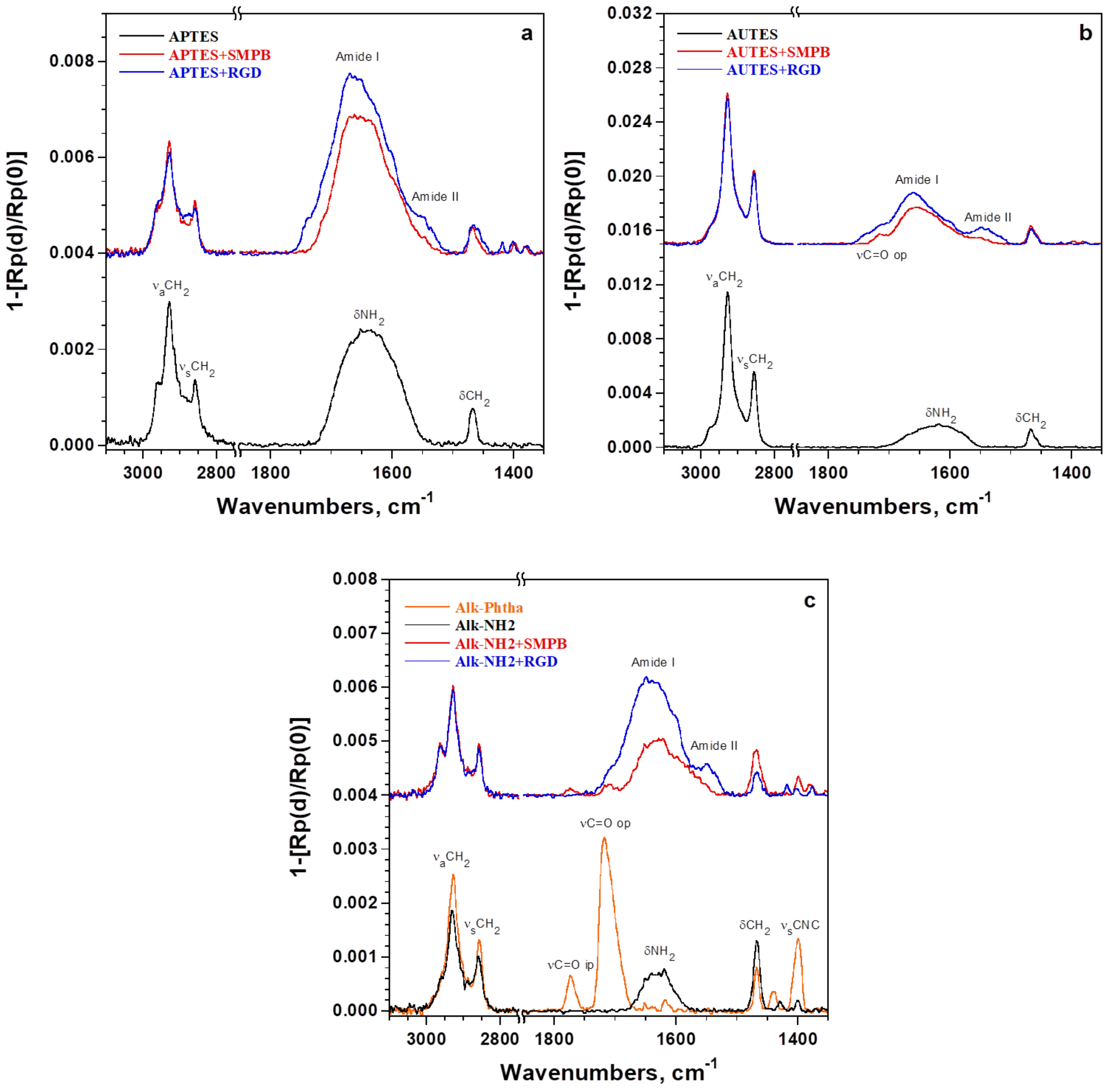

**Figure 2**. PM-IRRAS spectra after silanization with APTES (a), AUTES (b), and Alk-Phtha (c) onto $SiO_2$/Au substrates and after RGD peptide grafting. Silanization of APTES and Alk-Phtha was achieved through spin coating using a 2% v/v solution in absolute ethanol, while AUTES silanization was conducted via immersion in a 2% v/v solution in anhydrous toluene.

The three amino-terminated surfaces were immersed in the heterobifunctional SMPB crosslinker solution to afford the corresponding maleimide-terminated surface. The characteristic band at 1715 $cm^{-1}$ assigned to the out-of-phase carbonyl stretching vibrations of the maleimide moiety appeared clearly on the PM-IRRAS spectra of Alk-$NH_2$+SMPB and AUTES+SMPB[37,38]. In addition, the new amide bond was revealed by the presence of the corresponding amide I and amide II bands around 1650 $cm^{-1}$ and 1545 $cm^{-1}$, respectively, that confirm the covalent grafting of SMPB. For APTES+SMPB, the broad band of amino group centered at 1650 $cm^{-1}$ masks the corresponding bands of SMPB, which renders the extent of the reaction with SMPB difficult to evaluate. The cell-adhesive peptide RGD bearing a sulfydryl group was immobilized on the maleimide pendent surface through a stable covalent thioether bond. The PM-IRRAS spectra after the peptide grafting show that the intensities of amide I and amide II bands for Alk-$NH_2$+RGD increased significantly more than for APTES+RGD and AUTES+RGD.

## Silanization impacts cell adhesion

Silanization is a critical step in surface functionalization with protein mimetic peptides derived from ECM proteins[36]. Different silanization strategies were explored to assess the impact not only on the density of surface-immobilized peptides but also on their accessibility, bioactivity, and how this consequently impacts cell behavior. Cell adhesion tests were conducted on silicon surfaces functionalized with APTES+RGD, AUTES+RGD, and Alk-$NH_2$+RGD.

Cell morphology, dimensions, and arrangement are pivotal in optimally operating natural tissues and organs. In our investigation, we focused on quantifying the average area of individual cells (cell area, $\mu m^2$). Figure 3 illustrates how human bone marrow-derived mesenchymal stem cells (hBMSCs) exhibit diverse adhesion behaviors when adhering to our RGD-modified surfaces through various silane molecules and silanization techniques. Cells exhibited a pronounced spread morphology when cultured on surfaces functionalized with AUTES+RGD and Alk-$NH_2$+RGD. The characteristic elongation of cell bodies evidenced this. In contrast, cells appeared quiescent on control substrates with a rounded shape.

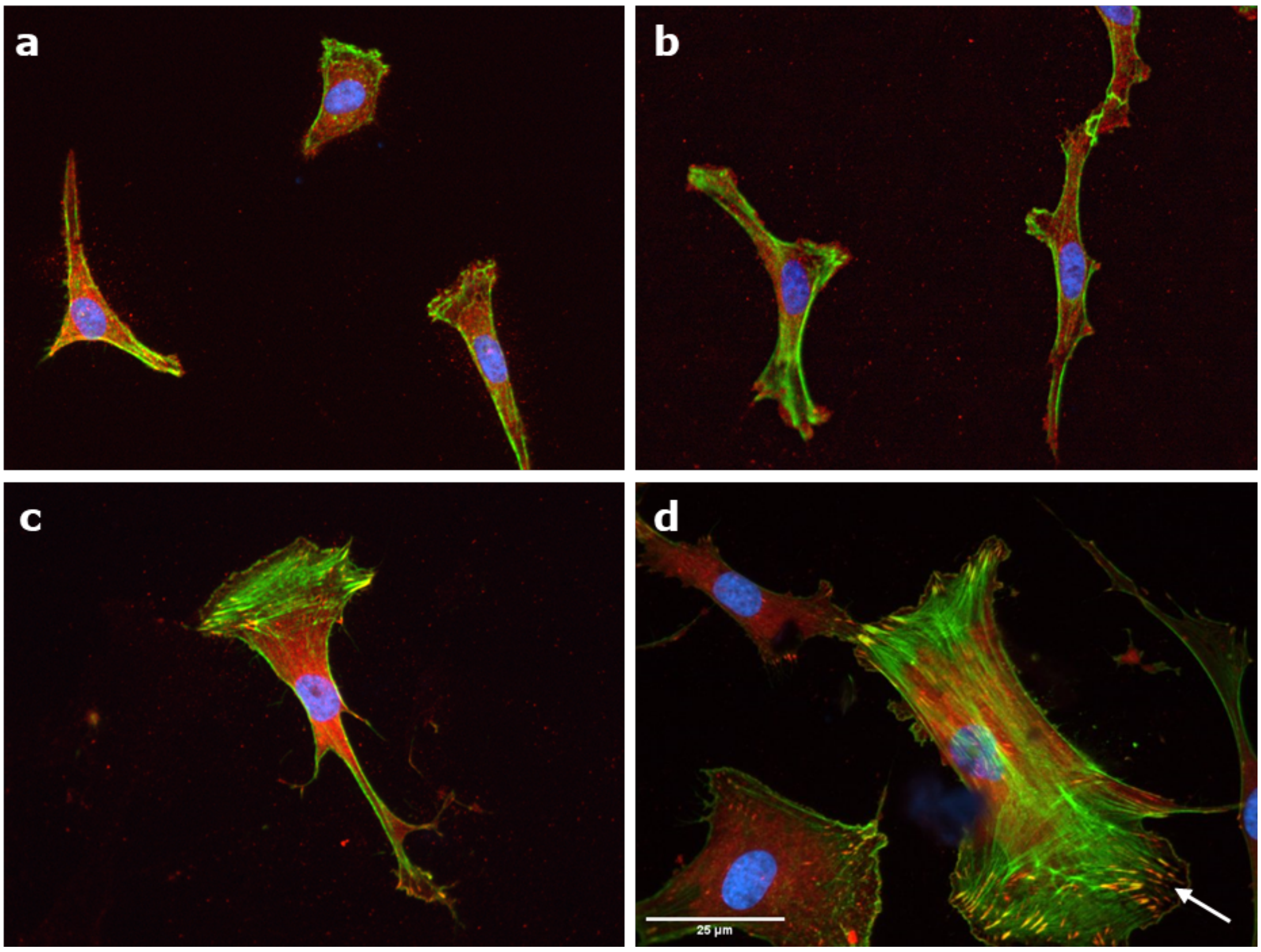

**Figure 3**. Representative fluorescence images of hBMSCs were seeded for 24 h in a DMEM medium on control (a), APTES+RGD (b), AUTES+RGD (c), and Alk-$NH_2$+RGD (d) substrates in serum-free conditions. Nucleus (Blue), F-actin (Green), and Vinculin (Red). The scale bar is 25 µm.

The measured cell area was 2566, 4069, 5051, and 5425 $\mu m^2$ for control, APTES+RGD, AUTES+RGD, and Alk-$NH_2$+RGD, respectively (Figure 4A). The cellular area of hBMSCs is significantly larger on all surfaces functionalized with RGD peptides compared to the control (virgin silicon). The cell area is lower after 24 h of culture for cells seeded on control and functionalized materials with APTES+RGD than those with AUTES+RGD and Alk-$NH_2$+RGD.

For this reason, F-actin and vinculin expressions were investigated to ascertain whether different silanized RGD-grafted surfaces may directly affect cytoskeletal changes and focal adhesion formation. Quantification of fluorescence intensity in F-actin showed that hBMSCs adherent to the Alk-$NH_2$+RGD substrate had a significantly higher fluorescence intensity than the other conditions, whose intensity was comparable to that found in the control samples (Figure 4B). Variations in the arrangement of the actin cytoskeleton were observed as well. hBMSCs cultured on surfaces treated with Alk-$NH_2$+RGD (Figure 3d) and AUTES+RGD substrates (Figure 3c) displayed large and well-organized actin stress fibers, whereas cells on the APTES+RGD surfaces exhibited a more disorganized actin cytoskeleton consisting of thin fibers and individual filaments extending outward from the cell's main body (Figure 3b). Lastly, hBMSCs cultivated on the control surfaces resulted in having a highly disorganized actin cytoskeleton (Figure 3a).

Next, we assessed the expression of vinculin, an essential element in forming focal adhesions crucial for cell adhesion and spreading. First, it was evident that vinculin exhibited varying distribution patterns across the different substrates. On both the control and APTES+RGD substrates, vinculin appeared prominently near the nucleus, with a dispersed presence throughout the cytoplasm. Conversely, on the AUTES+RGD and especially on the Alk-$NH_2$+RGD surfaces, vinculin expression is primarily localized to the cell peripheries, precisely at the tips of cytoplasmic filopodia, forming characteristic dash-like structures (white arrow Figure 3d).

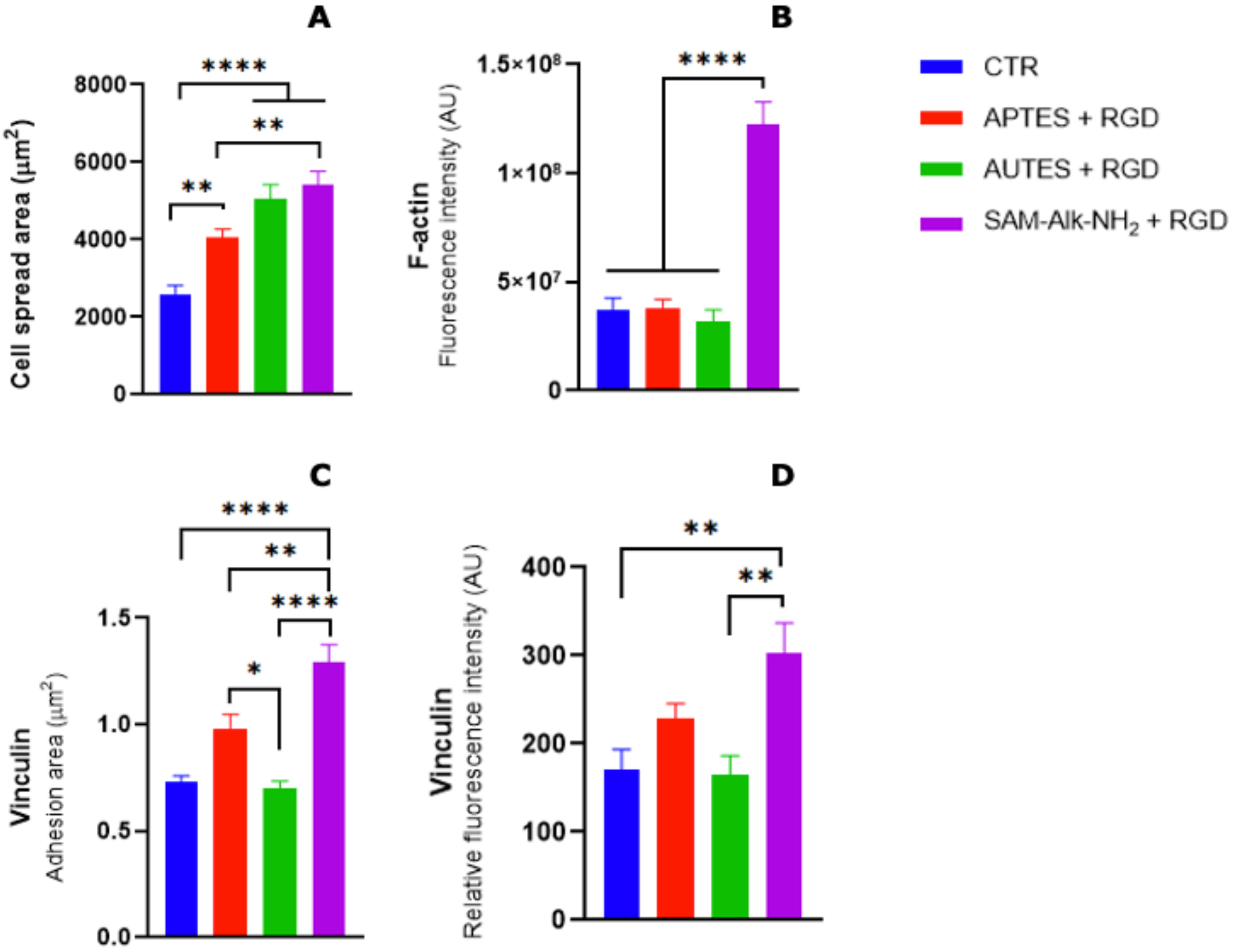

**Figure 4**. Histograms showing the change in cell spreading area of hBMSCs in response to RGD-substrates functionalized with different silanization strategies (A), F-actin expression (B), vinculin-containing focal adhesion area (C) and relative vinculin expression within focal adhesions (D). Statistical data are based on more than 80 cells in each group in two independent experiments (n > 80). Values are expressed as mean ± SEM. The statistically significant differences were determined by one-way ANOVA (**** $P < 0.0001$, *** $P < 0.001$, ** $P < 0.01$).

Therefore, the presence of focal adhesion points having an area within this range was assessed. Figure 4C shows that the substrate functionalized with Alk-$NH_2$+RGD had focal adhesion significantly more extensive than those that appeared on the control, APTES+RGD, and AUTES+RGD substrates. In addition, the fluorescence intensity in the focal adhesion sites was evaluated, revealing significantly elevated protein expression within the focal adhesion points of cells adhered to the Alk-$NH_2$+RGD, demonstrating a markedly higher vinculin expression compared to the other functionalized substrates (Figure 4D).

## 4. Discussion

The results presented in this study shed light on the critical role of silanization in surface functionalization with protein mimetic peptides derived from ECM proteins and its consequential impact on cell behavior. The comprehensive surface characterization aimed to elucidate the effects of silanization on the subsequent functionalization of biomaterials, specifically silicon wafers modified with different silane molecules and further grafted with the adhesive peptide RGD. An extensive understanding of the chemical and structural transformations at each surface modification stage was attained through the application of diverse analytical methodologies.

Most scientific literature reports APTES as the most used molecule for surface silanization to ensure subsequent grafting of biomolecules[39,40]. Consistent trends in elemental composition and surface distribution resulting from silanization through 3-aminopropyltriethoxysilane (APTES) were observed in X-ray photoelectron spectroscopy (XPS) analysis, aligning with existing literature[41]. However, a lack of reproducibility in APTES-based silanization was identified and attributed to the influence of numerous uncontrollable variables. To address this, we introduced an analogue molecule of APTES, 11-aminoundecyltriethoxysilane (AUTES), possessing an extended alkyl chain with 11 carbon atoms. Discrepancies in the existing literature regarding the efficacy of silanization methods prompted a comparative investigation by Liu et al. In their study, they explored the surface modification of silicon using APTES and AUTES. The findings revealed that AUTES-modified silicon surfaces effectively regulated ligand orientation, while APTES-modified surfaces exhibited heightened sensitivity in IgG adherence, particularly demonstrating a higher detection limit for anti-IgG. Consequently, APTES was identified as the more suitable candidate for biosensor applications[42]. However, Taylor et al. reported contrasting results, noting enhanced support for cell adhesion, proliferation, and viability on plain glass coverslips modified with AUTES, thereby introducing a conflicting perspective in the literature[43].

Previous studies have addressed the relationship between diverse chain lengths of aminosilanes and the behavior of human mesenchymal stem cells (hMSCs). Chen et al. carried out an investigation focused on elucidating the influence of $C_3$ and $C_{11}$ alkyl chain length in silane molecules, specifically assessing the modified deposition pattern of $NH_2$ groups on the substrate surface[41]. Their investigation resulted in the identification of the $C_{11}$ chain length to increase the roughness at the nanoscale, providing further osteoinductive properties of the modified material, which was lacking in the other chain lengths tested.

In a similar investigation, Taylor et al. observed that silanization with a longer carbon chain molecule ($C_{11}$) produced significantly superior results for neuronal cells in terms of adhesion and differentiation. In particular, the average length of neurites extending from primary neurons was significantly longer when grown on long-chain aminosilanes than when grown on short-chain aminosilanes ($C_3$). This study shows how the choice of a long-chain silane increases hydrophobicity and surface roughness, impacting cell differentiation[44].

The present study not only investigated the effects of different lengths ($C_3$ *vs* $C_{11}$) of silane chains, but also made a distinctive contribution by employing a silane molecule, Alk-Phtha, with the same chemical structure as AUTES but presenting a protecting N-phthalimide group on the terminal amine. This comparative study drew attention to a possible direct correlation between surface modification induced by silane molecules with a protecting group and subsequent alterations in cell behavior.

Indeed, XPS results obtained in the present study, showed a consistency with what is found in the literature concerning the chemical elemental distribution on the surface upon AUTES silanization[41,43]. However, XPS survey analysis revealed a different chemical composition after Alk-Phtha deposition, although after deprotection, the resulting molecule (Alk-$NH_2$) has the same chemical structure as AUTES. The observed differences were particularly pronounced in the high-resolution spectra, with AUTES showing a higher concentration on the surface than Alk-$NH_2$. The substantial differences observed in the high-resolution Si2p spectrum after the silanization process reveal pronounced variations that can be attributed to the distinctive molecular characteristics of the silanes. The hypothesis of multilayer formation during AUTES silanization was supported by the increased presence of C and N percentages, together with a decrease in the Si metal peak, which, on the other hand, in the case of monolayers should still be well detectable by XPS analysis. Indeed, the apparent correlation between the presence of free $NH_2$ groups in AUTES-modified substrates and the appearance of silanol bonds suggests potential multilayer formation explained by the observed changes in the Si spectrum.

This proposition was further validated by PM-IRRAS analysis. The intensity of the νaCH2 and νsCH2 bands at 2927 and 2855 cm-1, respectively, seems to be influenced by the total amount of silane molecules. Despite sharing the same chemical structure, AUTES and Alk-NH2 yielded distinct PM-IRRAS spectra. Notably, in the AUTES spectrum, the methylene band IRRAS intensity was significantly higher than that of the Alk-Phtha spectrum. This disparity may be attributed to the presence of a multilayer in the case of AUTES deposition. Conversely, when comparing APTES and Alk-Phtha modified surfaces, one would expect a lower contribution of the methylene bands in the APTES spectrum due to its alkyl chain consisting of only three carbon atoms, as opposed to the Alk-Phtha with 11 carbon atoms in its alkyl chain. Nevertheless, the two spectra displayed comparable peak intensities, suggesting that APTES deposition likely results in the creation of multilayers on the SiO2/Au surface. It is interesting to note that the general spectrum intensity of Alk-Phtha was consistent with the existence of the monolayer in the literature[24]. All in all, this study clearly shows the unfavorable impact of the terminal free amine group on the formation of a monolayer whatever the deposition method. The multiple possible interactions of the amine group during silanization lead to a disorganized film and therefore favor 3D polymerization (multilayers). Conversely, when Alk-Phtha interacts with the surface, after hydrolysis only trisilanol can establish H-bonding interactions with the silanols groups of the silica surface, enabling molecule orientation and 2D polymerization (monolayer). In addition, the analyses using XPS and PM-IRRAS techniques have provided consistent results, demonstrating the presence of either monolayers or multilayers depending on deposition conditions and silanes utilized. Accordingly, the complexity of assessing the density of accessible amine functions on the surface may be difficult due to the intricate nature of molecular compactness of the layers.

Interestingly, despite the seemingly lower concentration on the surface, Alk-$NH_2$ demonstrated potential advantages, likely attributable to the protecting phthalimide group. This group, shielding the $NH_2$ moiety, facilitated a more homogeneous distribution on the surface. The subsequent deprotection of the phthalimide group, confirmed by PM-IRRAS, resulted in enhanced accessibility for peptide grafting. These results clearly show the advantage of using the phthalimide protecting group as a strategy to prepare amino-terminated SAMs, which allows more accessible amino groups on the surface, unlike commercial aminosilanes which led to disordered layers with unavailable amines engaged in multiple interactions. These findings emphasize the critical role of functional group protection and molecular organization in controlling surface reactions during silanization[35]. Silane molecules have conventionally served as agents for silanization through traditional immersion or vapor deposition techniques. In this investigation, a novel approach was presented: silanization employing spin coating, which, to the best of our knowledge, is the first time it has been used for subsequent grafting of biomolecules. Prior research has established spin coating as an effective silanization method, comparable to conventional approaches[27]. The distinctive advantages of this methodology are twofold. Firstly, it facilitates the use of fewer solvents, thereby mitigating the generation of chemical waste that would otherwise ensue with methods such as immersion silanization. Simultaneously, it substantially diminishes the duration of the silanization process, enhancing productivity within a shorter timeframe, a pivotal consideration, particularly in the context of industrial-scale processes.
A cell adhesion study was carried out, where hBMSCs cultured on differently silanized materials yet grafted with the same RGD peptide displayed distinct behavior. The observed variations in cell spreading led to exploring cytoskeletal changes and focal adhesion formation. Our findings align with previous studies observations on the impact of cell shape alterations on focal adhesion (FA) organization and orientation relative to cytoskeletal structures[45–51]. Specifically, mesenchymal stem cells (MSCs) respond to extracellular cues by forming cellular adhesive organelles (FAs)[52], connecting with radial stress fibers[53], and bridging with transverse fibers to establish a dynamic cytoskeletal network, indicative of early commitment[54]. Our investigation expanded on these dynamics by noting significant differences in F-actin expression and organization across various substrates. Cells adhering to control and APTES+RGD substrates demonstrated reduced cell area, correlating with diminished F-actin fluorescence intensity. Conversely, cells cultured on Alk-$NH_2$+RGD displayed larger cell areas and notably higher fluorescence intensity in relation to F-actin protein expression. This might imply a correlation between cell area and cytoskeletal rearrangement, as evidenced by altered F-actin expression levels. Focal adhesion analysis further supported these findings. These formations represent structured clusters of specialized proteins, such as cytoskeletal, signalling, and regulatory proteins, distributed along the basal surface of adherent cells[10,55]. They establish physical connections between the extracellular matrix and the actin cytoskeleton through transmembrane receptor integrins, facilitating signal transduction between the cell and its environment [56–58]. Based on a previous study[46,54], emphasis was placed on vinculin-containing focal adhesions formed within 24 h with a total area ranging from 0.5 – 5 $\mu m^2$. The focal adhesion sites area and fluorescence intensity were notably higher for Alk-$NH_2$+RGD. This suggests that the silanization method plays a crucial role in the accessibility of $NH_2$ groups and therefore how the peptides are distributed onto the surface. Alk-$NH_2$+RGD facilitated a substantially more significant expansion of adhesion site area and vinculin protein expression within focal adhesion sites. These findings contribute valuable insights into the interplay between substrate-induced alterations in cell morphology and associated cytoskeletal dynamics.

In the context of biomaterial surface functionalization, APTES in immersion has been widely recognized as a well-established and effective silanization method, often considered the first choice due to its ability to allow stable grafting of biomolecules by creating stable and high-density covalent bonds on the surface[59–61]. However, the density of molecules on the surface does not necessarily correlate directly with effectiveness. Indeed, a higher density may not necessarily lead to increased bioactivity following bioactive molecule grafting[62]. Aligned with these findings, it was demonstrated that a lower density on the surface, confirmed by XPS analysis, yielded the most favorable results in cell adhesion. To further explore and compare with the conventional APTES with a short alkyl chain ($C_3$), two other molecules were introduced, AUTES, with a longer alkyl chain ($C_{11}$), and Alk-Phtha, capable of creating a self-assembled monolayer with a steric hindrance effect due to the N-protecting phthalimide group. Exploration of the standard silanization methods was also undertaken. While widely used, solution-based reactions present challenges such as substantial solvent use and chemical waste generation. On the other hand, vapor-phase deposition eliminates the need for organic solvents but requires ultra-high vacuum conditions and elevated temperatures[27,63]. A novel approach was introduced with this work, by implementing spin coating as a silanization method, a technique utilized for other purposes[64] but was never used to immobilized biomolecules. Surprisingly, our comparison revealed that, once deprotected, Alk-$NH_2$ exhibited the best cell adhesion behavior, potentially due to better accessibility of the amino groups, even at a lower density, allowing for improved grafting and enhanced interaction with cell receptors.

The present study challenges the conventional silanization techniques and introduces spin coating as a versatile and effective alternative. Different silanization processes provided a further distribution of the peptide that has been perceived differently by cells, which reacted very differently to the same RGD peptide grafted on the surface just because it was grafted on different silanized surfaces. This study demonstrated that silanization constitutes a critical step in the functionalization of biomaterials, as evidenced by the cell adhesion study. The molecule Alk-Phtha, emerged as a promising candidate for achieving better grafting orientation and enhanced accessibility for peptide grafting.

This work introduces a novel method and molecule for biological purposes, challenging conventional approaches and demonstrating that comparable, if not superior, results can be attained through a simpler, more cost-effective, and environmentally sustainable methodology. This novel perspective on biomaterial surface functionalization opens the door to re-evaluating the initial layer of modification, which plays a pivotal role in determining the success of subsequent layers and overall outcomes. These results collectively validate the sequential and successful alteration of surface properties, showing the intricate control achieved in tailoring surface properties. The study demonstrates that the functionalization strategy is very important and can determine the quality of functionalization in terms of accessibility of terminal functions and thus responsiveness and homogeneity, as well as having a strong influence on cells. This deviation from conventional protocols gives researchers greater autonomy in modeling biomaterial fabrication processes, paving the way for the creation of more efficient and customized techniques to address biomedical challenges.

## Conclusions

In conclusion, this study provides valuable insights into the interplay between substrate-induced alterations in cell morphology and associated cytoskeletal dynamics. We challenge conventional techniques by exploring novel silanization methodologies and introducing silanization by spin coating as a versatile and effective alternative for grafting bioactive molecules onto the surface. The functionalization strategy using a protected aminosilane provides a better controlled surface functionalization, resulting in a homogeneous SAM. The findings demonstrate that different silanization processes result in varied peptide distributions, leading to distinct cellular responses. Silanization emerges as a critical step in the functionalization of biomaterials, influencing surface characteristics that, in turn, impact cell adhesion. The sequential alteration of surface properties showcases the intricate control achieved in tailoring biomaterial surfaces.

This deviation from conventional protocols allows for refining surface engineering methodologies and achieving tailored surface properties, offering increased autonomy in shaping biomaterial fabrication processes for addressing diverse biomedical challenges. This research highlights the fundamental importance of silanization as a foundational stage, unveiling its significant impact on the subsequent functionality of biomaterials and the effectiveness of interactions at the cell-surface interface.

*ACKNOWLEDGEMENTS*

The authors thank Anthony Vial and Michael Molinari from the VibrAFM platform at CBMN for the AFM experiments. This work was supported by the Natural Science and Engineering Research Council of Canada (G.L), as well as the Centre Québécois sur les Matériaux Fonctionnels (G.L). M.K is the recipient of a scholarship from the Fondation du CHU de Québec-Université Laval. M.C.D would like to thank the French National Research Agency (ANR-21-CE06-0031-02) for its support, as well as the Nouvelle Aquitaine Region (contract n° 2017-1R30109-00013176) and the Department of Health Sciences and Technologies from the University of Bordeaux for their support.